\documentclass[a4paper,fleqn]{cas-dc}

\usepackage[authoryear,round]{natbib}
\usepackage{url}
\usepackage{algorithmic}
\usepackage{graphicx}
\usepackage[linesnumbered, ruled,vlined]{algorithm2e}
\usepackage{enumitem}
\usepackage{color}
\usepackage{stfloats}
\usepackage{subcaption}
\usepackage{multirow}
\usepackage{flushend}
\usepackage{bbding}
\usepackage{ragged2e}
\usepackage{footnote}
\usepackage{color}
\usepackage{float}

\begin{document}
	\begin{sloppypar}
		\let\WriteBookmarks\relax
		\def\floatpagepagefraction{1}
		\def\textpagefraction{.001}
		\shorttitle{Hierarchical feature extraction model for ASD classification}   
		\shortauthors{Yiqian Luo et~al.}
		
	  \title [mode = title]{Hierarchical feature extraction on functional brain networks for autism spectrum disorder identification with resting-state fMRI data}
		
		\author[1]{Yiqian Luo}

		\credit{Methodology, Software, Validation, Writing-Original Draft, Writing-Review and Editing, Visualization}	
		
		\author[1]{Qiurong Chen}
		\credit{Software,Writing-Review and Editing}	
		
		\author[2]{Fali Li}
		\credit{Methodology, Writing-Review and Editing}	

		\author[3, 4]{Liang Yi}
		\credit{Writing-Review and Editing}	

		\author[1,2]{Peng Xu}
		\credit{Conceptualization, Methodology, Writing-Review and Editing}	
		\cormark[1]
		\ead{xupeng@uestc.edu.cn}

		\author[1,2]{Yangsong Zhang}
		\cormark[1]
		\ead{zhangysacademy@gmail.com}
		\credit{Supervision, Conceptualization, Methodology, Writing-Review and Editing}

		
		
		\address[1]{Laboratory for Brain Science and Artificial Intelligence, School of Computer Science and Technology, Southwest University of Science and Technology, Mianyang, China}
		\address[2]{MOE Key Laboratory for NeuroInformation, Clinical Hospital of Chengdu Brain Science Institute, and Center for Information in BioMedicine, School of Life Science and Technology, University of Electronic Science and Technology of China, Chengdu, China}
		\address[3]{Department of Neurology, Sichuan Provincial People's Hospital, University of Electronic Science and Technology of China, Chengdu, China}
		\address[4]{Chinese Academy of Sciences Sichuan Translational Medicine Research Hospital, Chengdu, China}

		\cortext[cor1]{Corresponding author}

\begin{abstract}
Autism Spectrum Disorder (ASD) is a pervasive developmental disorder of the central nervous system, primarily manifesting in childhood. It is characterized by atypical and repetitive behaviors. Currently, diagnostic methods mainly rely on questionnaire surveys and behavioral observations, which are prone to misdiagnosis due to their subjective nature. With advancements in medical imaging, MR imaging-based diagnostics have emerged as a more objective alternative. In this paper, we propose a Hierarchical Neural Network model for ASD identification, termed ASD-HNet, which hierarchically extracts features from functional brain networks based on resting-state functional magnetic resonance imaging (rs-fMRI) data. This hierarchical approach enhances the extraction of brain representations, improving diagnostic accuracy and aiding in the identification of brain regions associated with ASD. Specifically, features are extracted at three levels: (1) the local region of interest (ROI) scale, (2) the community scale, and (3) the global representation scale. At the ROI scale, graph convolution is employed to transfer features between ROIs. At the community scale, functional gradients are introduced, and a K-Means clustering algorithm is applied to group ROIs with similar functional gradients into communities. Features from ROIs within the same community are then extracted to characterize the communities. At the global representation scale, we extract global features from the whole community-scale brain networks to represent the entire brain. We validate the effectiveness of our method using the publicly available Autism Brain Imaging Data Exchange I (ABIDE-I) dataset. Experimental results demonstrate that ASD-HNet outperforms existing methods. The code is available at https://github.com/LYQbyte/ASD-HNet.
\end{abstract}


\maketitle

\section{Introduction}
Autism spectrum disorder (ASD) is a type of pervasive developmental disorder affecting the central nervous system~\citep{pandolfi2018screening}, approximately 1\% of the world's population is affected by ASD. Moreover, ASD is associated with a high incidence of concurrent disorders, such as depression and anxiety~\citep{vohra2017comorbidity}. Autism spectrum disorder typically manifests in childhood, with its associated impairments often persisting throughout an individual's lifetime. Many individuals with ASD require lifelong support to manage their condition~\citep{lord2018autism}. ASD is characterized by a broad spectrum of symptoms, including deficits in social communication and the presence of restrictive, repetitive behaviors~\citep{hirota2023autism, bhat2014automated}. The disorder poses significant challenges to both the physical and mental health of affected individuals, profoundly impacting their quality of life as well as that of their families. Despite its profound effects, the precise pathogenesis of ASD remains elusive, making accurate diagnosis particularly challenging.
	
Unfortunately, there are no definitive drugs and methods for the treatment of ASD, and the pathological mechanisms of ASD remain for further exploring~\citep{geschwind2009advances}. At present, the most effective strategies involve providing advanced behavioral training for individuals with ASD, implementing early intervention programs, and addressing problematic behaviors associated with the condition. These approaches underscore the critical importance of early and accurate diagnosis of ASD. However, relying solely on the manifestation of behavioral symptoms often results in delayed diagnosis, potentially causing patients to miss the optimal window for intervention and complicating treatment efforts. Currently, the majority of diagnostic methods rely on questionnaires and behavioral observations to determine whether an individual has ASD~\citep{american2013diagnostic}. These approaches are highly subjective and susceptible to environmental influences, which increases the risk of misdiagnosis~\citep{wang2022multi}. Consequently, there is an urgent need to develop highly accurate, pathology-based diagnostic tools and to identify reliable biomarkers associated with ASD to improve diagnostic precision and facilitate timely intervention.

Benefiting from the rapid development of medical image technology, imaging-based approaches provides us with important tools to analyze and study neurophysiology. Functional magnetic resonance imaging (fMRI), as a non-invasive neuroimaging technique~\citep{matthews2004functional}, can measure the blood oxygen level dependent (BOLD) changes caused by neuronal activity throughout the brain at a series of time points~\citep{friston1994analysis}. Numerous studies have demonstrated the feasibility of utilizing resting-state functional magnetic resonance imaging (rs-fMRI) to investigate the interactions between regions of interest (ROIs) in the brains of individuals with mental disorders~\citep{dvornek2018combining}. In recent years, a growing body of research has integrated deep learning techniques with rs-fMRI to explore and analyze mental illnesses, achieving notable progress~\citep{dong2020compression, luppi2022synergistic}. In most studies, the brain is partitioned into multiple ROIs using predefined brain atlases, followed by the extraction of BOLD time series from these ROIs. Subsequently, the correlations between the BOLD time series of different ROIs are computed to construct brain functional connectivity networks (FCNs), which are then used for further analysis and exploration of brain function~\citep{dong2024brain, wang2023covariance, hong2019atypical}.
	
The brain FCN extracted from resting-state fMRI data can reflect the functional interactions between brain regions, which serves as a very effective representation tool to explore the normal and altered brain function~\citep{van2010exploring}. At present, numerous deep learning methods utilize brain FCNs to implement ASD identification. For instance, Eslami et al. proposed using a data-driven linear approach to enhance the training data and using autoencoders to achieve the diagnosis of ASD~\citep{eslami2019asd}. Kawahara et al. proposed a network architecture called BrainNetCNN, which has special edge-to-edge, edge-to-node, and node-to-graph convolutional filters, and takes into account the non-Euclidean nature of brain networks~\citep{kawahara2017brainnetcnn}. Jiang et al. proposed a graph network model Hi-GCN, which first extracts the features of each subject using the ROI-level graph network, and then constructs a population graph network with the extracted features of each subject as nodes~\citep{jiang2020hi}. Hi-GCN takes into account the spatial properties of brain networks, as well as the structure of the network directly between people. Recently, the DGAIB model proposed by Dong et al.~\citep{dong2024brain} has utilized graph neural networks to carry out effective feature transmission in brain regions, combining an information bottleneck to filter out irrelevant information related to the target task, thereby applying the retained effective features to improve classification performance. Dong et al. proposed to build an individualized and a common functional connectivity networks, and then use the proposed multiview brain Transformer module to carry out parallel feature extraction learning for the two FCNs~\citep{dong2024multiview}. Finally, the characteristics of the two branches are fused to make the final diagnosis classification. 

In recent years, there has been some progress in the research of constructing multi-scale brain network to realize the diagnosis and classification of ASD. For instance, Liu et al. used a set of atlases for multiscale brain parcellation, and then progressively extracted functional connectivity networks from small scale to large scale by constructing mapping relationships between different scale brain maps~\citep{liu2023hierarchical}. Similarly, Bannadabhavi et al. proposed a community-aware approach by using two different scales of altas and then constructing a map of ROIs to communities based on the division of the two brain atlas~\citep{bannadabhavi2023community}. Then transformer~\citep{vaswani2017attention} is used to extract ROI-scale network into community-scale brain network for corresponding diagnosis and classification. However, the existing methods all need to manually divide the mapping relationship between ROIs and communities, which is very complicated and cumbersome.

Previous deep learning models often lack the model structure exploration of physiological mechanisms and fail to effectively mine multi-scale features, ranging from local regions to the whole-brain scale, which ultimately leads to suboptimal diagnostic performance. In the current study, we combine resting-state functional magnetic resonance imaging with deep learning techniques to design a network model from a brain physiology perspective, with the goal of improving ASD detection performance. We propose a hybrid neural network model for ASD identification, termed ASD-HNet, which hierarchically extracts features from functional brain networks at three scales: local ROI scale, community scale, and global representation scale. Specifically, we use the Automated Anatomical Labeling (AAL) template~\citep{tzourio2002automated} to divide the brain into $N_R$
($N_R$=116) regions of interest, extract the BOLD signals from each ROI, and compute the functional connectivity network. A graph convolutional network (GCN)~\citep{niepert2016learning} is then employed to transfer information between ROIs. Next, we introduce functional gradients to automatically cluster ROIs into communities, as opposed to relying on pre-defined atlases. This clustering approach groups ROIs with similar functional patterns, allowing us to extract community-scale features. Finally, global brain convolution is applied to the community-scale features to integrate information at the global scale. Additionally, prototype learning is incorporated for classification based on the features extracted by ASD-HNet, leading to improved identification accuracy. Experiments on the publicly available Autism Brain Imaging Data Exchange I (ABIDE-I) dataset validate the superior performance of the proposed ASD-HNet. Moreover, extensive experiments on model generalization further demonstrate the robustness and adaptability of our approach across different datasets.

	%
		
		
		
		

\section{Materials and Methodology}
\subsection{Functional Gradients}

Functional gradients in the brain offer a more comprehensive perspective by capturing the continuous spatial patterns of connectivity beyond the segregated networks typically observed in the human brain~\citep{guell2018functional, huntenburg2018large}. It's a gradient-based approach, which performs a nonlinear decomposition of high-dimensional resting-state functional connectivity (rs-FC), can identify distinct brain functional levels by representing connectivity in a continuous, low-dimensional space~\citep{dong2020compression}. Functional gradients have been extensively applied in brain analysis~\citep{dong2020compression, guo2023functional, gong2023connectivity}. In the context of various mental disorders, functional gradients provide a clear indication of abnormal functional hierarchies across brain regions.

The functional gradients were also used to study the ASD. For instance, Hong et al.~\citep{hong2019atypical} found functional gradients disorder in autistic patients, which showed that the functional distances between the transmodal regions and the unimodal regions decreased, that is, the functional independence of the brain region was reduced. Urchs et al. used ABIDE-I dataset~\citep{di2014autism} to carry out corresponding functional gradients analysis for autism, and also revealed the compression of gradients (less functional separation) in ASD~\citep{urchs2022functional}. Although functional gradients has been used in data analysis for ASD, combining functional gradients with deep learning for classification and diagnosis of ASD remain to be explored. In this paper, we proposed a deep learning model by leveraging functional gradients to achieve better ASD identification performance.
	
\begin{figure}[!htbp]
		\centering
		\includegraphics[width=\linewidth]{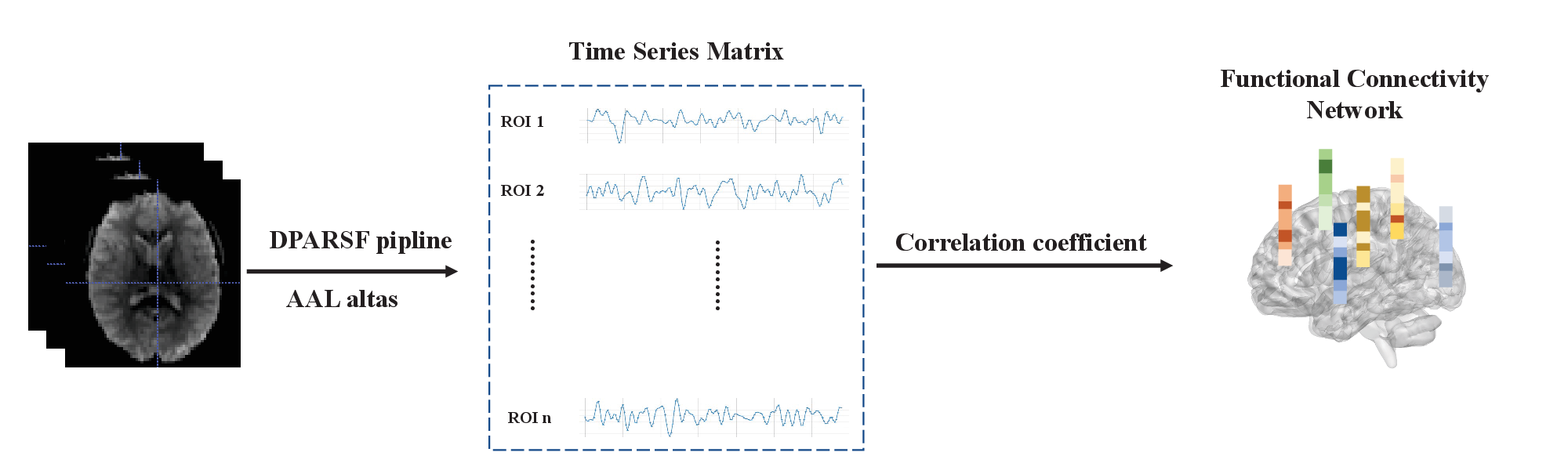}
		\caption{The pipelines process and computation of functional connectivity network.}
		\label{pipeline_fig}
\end{figure}
	
\subsection{Materials and data preprocessing}
To verify the performance of the proposed method, we conducted the experiment on the publicly available ABIDE-I dataset, including rs-fMRI and phenotypic data of 1112 subjects from 17 different sites~\citep{di2014autism,craddock2013neuro}. In current study, we utilized rs-fMRI data from 848 subjects, including 396 patients with ASD and 452 health controls (HCs). The demographic information of the studied subjects is reported in Table~\ref{Dataset_details}.

\begin{table}[!htbp]
	\caption{Demographic information of subjects from ABIDE-I dataset.}
	\renewcommand\arraystretch{1.2}
	\tabcolsep=0.1cm
	\label{Dataset_details}
	\begin{tabular}{l|cccc}
		\hline
		Dataset                & Category & Subject & Gender(M/F) & Age(Mean $\pm$ std) \\ \hline
		\multirow{2}{*}{ABIDE-I} & ASD      & 396     & 347/49      & 17.78 $\pm$ 9.01    \\
		& HC       & 452     & 371/81      & 16.75  $\pm$ 7.25    \\ \hline
	\end{tabular}
\end{table}
	
For data preprocessing, we used the data Processing Assistant for Resting-State fMRI (DPARSF) to preprocess rs-fMRI data~\citep{yan2010dparsf}. The specific steps are as follows: (1) Discarding the first 5 volumes. (2) Correcting of slice-related delay. (3) Correcting head movement. (4) Using EPI template normalization in MNI space, resampling to 3 mm resolution of $3\times3\times3$. (5) Smoothing with a 4 mm full-width at half-maximum (FWHM) Gaussian kernel. (6) Linear detrending and temporal band-pass filtering of BOLD signal (0.01-0.10 Hz). (7) Harmful signals that return head movement parameters, white matter, cerebrospinal fluid (CSF), and global signals. The registered fMRI volumes are then partitioned into $N_R$ ROIs using the AAL template, and BOLD time series are obtained for each ROI. Based on these time series of all the ROIs, we obtained the brain FCN matrix by calculating the Pearson correlation coefficient with the formula~(\ref{PCC}).

	\begin{equation}
		{\rho _{ij}} = \frac{{\sum\nolimits_{t = 1}^T {\left( {{i_t} - \overline i } \right)\left( {{j_t} - \overline j } \right)} }}{{\sqrt {\sum\nolimits_{t = 1}^T {{{\left( {{i_t} - \overline i } \right)}^2}} } \sqrt {\sum\nolimits_{t = 1}^T {{{\left( {{j_t} - \overline j } \right)}^2}} } }}
		\label{PCC}
	\end{equation}
	
	
Here $i_t$ and $j_t$ ($t=1,2,\cdots,T$) represent the time series of  $i$-th and  $j$-th brain regions, and the $\overline i$ and $\overline j$ are the average of two time series of brain region $i$ and $j$, respectively. $T$ represents the total length of time series.

	\begin{figure*}
		\centering
		\includegraphics[width=\linewidth]{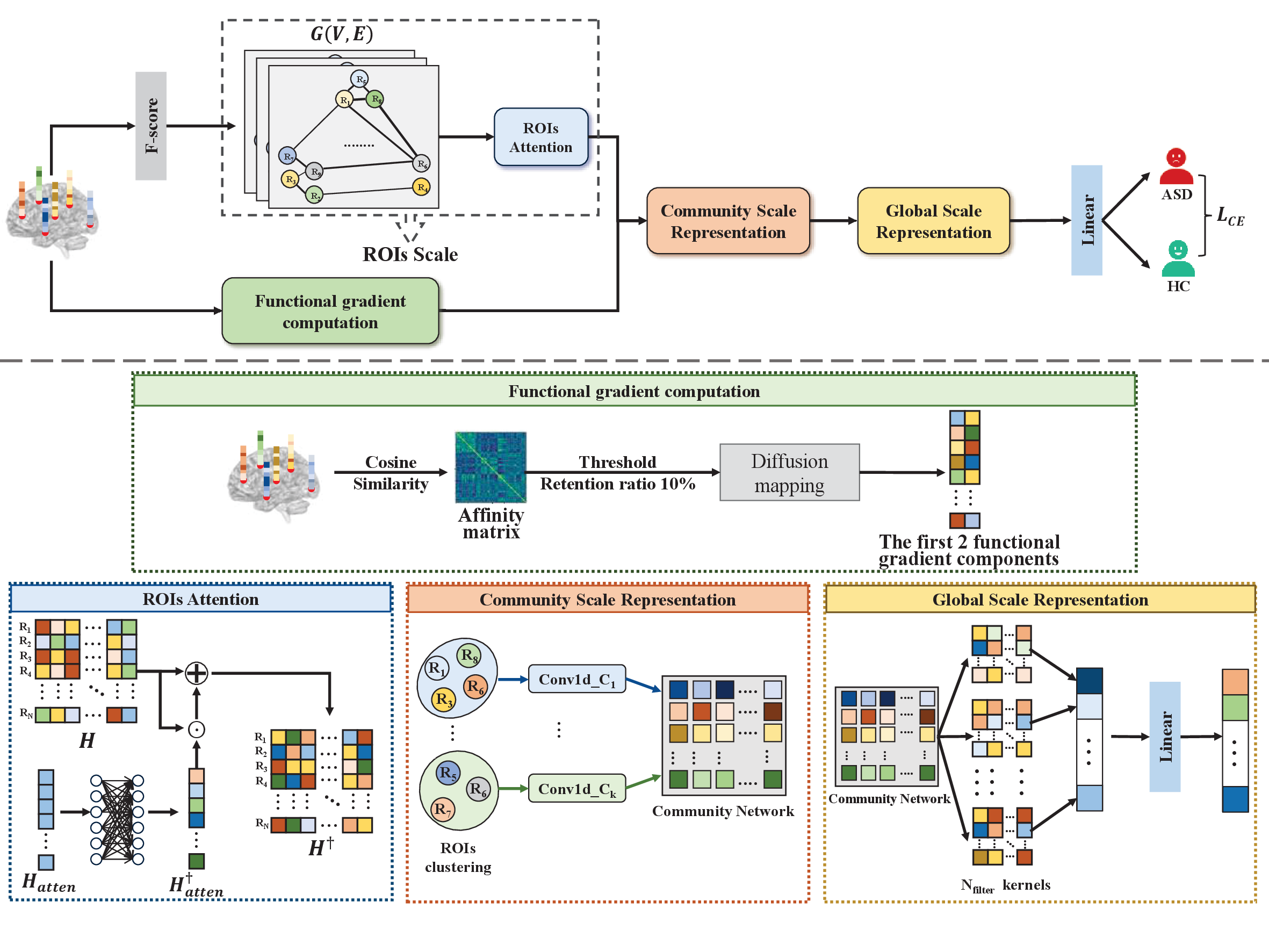}
		\caption{The diagram of the proposed hybrid neural network model for ASD identification (ASD-HNet).}
		\label{model_fig}
	\end{figure*}

\subsection{Proposed Method}
The schematic diagram of the proposed ASD-HNet is shown in Figure~\ref{model_fig}. This model is designed to progressively extract the discriminative and robust representations based on FCNs from three different scales, i.e., local ROI scale, community scale, and global representation scale, to achieve ASD identification. More details are described below.

\subsubsection{The local ROI scale}

The suspicious connections calculated with the formula~(\ref{PCC}) could exist in the FCN, and they can degrade the classification performance of the model using raw FCN as input. To eliminate the influence, we adopted the $F$-score~\citep{chen2006combining} as a global feature selector to screen the features in the FCN. For each subject, since their FCN is a real symmetric matrix, we utilized the upper or lower triangle elements of the FCN matrix as the screened features which are then unfolded to a vector. Subsequently, we computed the $F$-score for each element in the vector by the formula~(\ref{F-score}).
	
	\begin{equation}
		\resizebox{.95\hsize}{!}{$
		{F_i} = \frac{{{{\left( {\overline {x_i^{ASD}}  - \overline {{x_i}} } \right)}^2} + {{\left( {\overline {x_i^{HC}}  - \overline {{x_i}} } \right)}^2}}}{{\frac{1}{{{N_{ASD}} - 1}}\sum\limits_{k = 1}^{{N_{ASD}}} {{{\left( {x_{k,i}^{ASD} - \overline {x_i^{ASD}} } \right)}^2}}  + \frac{1}{{{N_{HC}} - 1}}\sum\limits_{k = 1}^{{N_{HC}}} {{{\left( {x_{k,i}^{HC} - \overline {x_i^{HC}} } \right)}^2}} }}$}
		\label{F-score}
	\end{equation}
where $\overline {{x_i}}$, $\overline {x_i^{ASD}}$, $\overline {x_i^{HC}}$ represent the grand average of the $i$-th feature of all subjects, the group average of $i$-th feature of ASD subjects and the $i$-th feature of HC subjects, respectively. ${N_{ASD}}$ and ${N_{HC}}$ denote the number of  ASD subjects and HC subjects. The ${x_{k,i}^{ASD}}$ and ${x_{k,i}^{HC}}$ refers to the $i$-th feature of the $k$-th ASD subject and $k$-th HC subject, respectively.  
	

Using the F-score, we derived two sparse FCN matrices. Following the methods in~\citep{zhang2023detection,tong2023fmri}, we retained the top 10\% of connections in the functional connectivity matrix to construct a sparse FCN matrix, which served as the adjacency matrix ($E$) for building a graph network. Additionally, to mitigate noise in the node features, we further sparsified the FCN matrix by retaining a proportion ${\lambda _R}$ (${\lambda _R}$=50\%) based on the F-score criterion.


Based on these two sparse FCN matrices, we constructed a ROI-scale graph network $G(V,E)$, where $V$ is the node feature matrix and $E$ is the adjacency matrix. Then, we used graph convolutional neural network to conduct feature extraction on the ROI scale. Here, we used one layer of spatial-based GCN to complete the message transfer between ROIs, the specific implementation is as follows:

	
	\begin{equation}
		H = \sigma \left( {E \cdot V  \cdot W} \right)
		\label{GCN}
	\end{equation}
where $W$ is a layer-specific trainable weight matrix with the shape ($N_R$, $D_1$) ($D_1$=64) and $\sigma$ is the activation function. In order to make nodes pay attention to their own characteristics, the diagonal elements of the adjacency matrix $E$ are all set to 1.0. 
	
Following the graph convolution operation, we further employed a ROIs attention mechanism to weight each ROI features. Sepcifically, an attention weight matrix $H_{atten}$ was initialized with all elements set to 1.0. Subsequently, two successive fully connected layers were employed to learn the attention weights for each ROI. The operation of a single linear layer is shown in the following formula~(\ref{atten}). It should be noted that since ROIs are weighted using positive numbers, so the activation function ($\sigma_{atten}$) was ReLU in the fully connected layer.
	
	\begin{equation}
		H_{atten}^\dag  = \sigma_{atten} \left( {W_{atten}^\dag {H_{atten}} + Bias_{atten}^\dag } \right)
		\label{atten}
	\end{equation}

After obtaining the attention weights of ROIs, we applied these weights to multiply the features of each ROI. Subsequently, a residual connection was employed to integrate the weighted ROI features with the original ROI features, followed by a batch normalization operation, as illustrated in formula~(\ref{atten_roi}).
	
	\begin{equation}
		{H^\dag } = BN\left( {H_{atten}^\dag  \odot H + H} \right)
		\label{atten_roi}
	\end{equation}
where $BN$ denotes the BatchNorm1d operation and $\odot$ represents the dot product operation. 
 	
\subsubsection{The community scale}
The brain's community network comprises groups of functionally similar ROIs, each with a high degree of functional concentration. Community networks are crucial for understanding the functional organization of the brain~\citep{van2009functionally}. Functional communities have been shown to be relevant to understanding cognitive behavior~\citep{van2010exploring}, mental states~\citep{geerligs2015state}, and neurological and psychiatric disorders \citep{canario2021review}. We employed the community scale representation module to further extracted functional features at community-clustering scale as illustrated in Figure~\ref{model_fig}. Since functional gradients can well reflect the functional similarity between ROIs, we proposed to use functional gradients to automatically cluster ROIs into community networks.
	
We utilized whole group-level (include ASD subjects and HC subjects) average functional connectivity matrix inputs into BrainSpace~\citep{vos2020brainspace} to obtain group-level functional gradients. In general, the computation of functional gradients involves two steps. The first step is selecting a kernel function to compute the affinity matrix. In the second step, the resulting affinity matrix is subjected to nonlinear dimensionality reduction to obtain the functional gradients. In our work, we chose cosine similarity as the kernel function to compute the affinity matrix ($A(i,j)$). The calculation is expressed as follows:

	
	\begin{equation}
		A\left( {i,j} \right) = \cos sim\left( {{R_i},{R_j}} \right) = \frac{{{R_i}R_j^{Tr}}}{{\left\| {{R_i}} \right\|\left\| {{R_j}} \right\|}}
		\label{cossim}
	\end{equation}
where $R_i$ and $R_j$ represent the feature vectors of the $i$-th and $j$-th ROI in the FCN, respectively. And $\left\|  \cdot \right\|$ denotes the $l_2$-norm, $Tr$ stands for transpose operation. Before performing the nonlinear dimensionality reduction on the affinity matrix, we first apply sparsification to the matrix. Specifically, we retain only the top 10\%~\citep{gong2023connectivity,vos2020brainspace} of the strongest functional connections for each ROI, thereby extracting the most significant connections.
	
For nonlinear dimensionality reduction, we employed the diffusion mapping technique in current study. We calculated the diffusion matrix using the previously obtained affinity matrix $A(i,j)$. The specific calculation process is as follows.
	\begin{equation}
		{P_\beta } = D_\beta ^{ - 1}{W_\beta }
		\label{DM}
	\end{equation}
where $W_\beta$ is defined as formula~(\ref{DM_kernel}), $D_\beta$ is the degree matrix derived from $W_\beta$.
		\begin{equation}
		{W_\beta } = {D^{ - 1/\beta }}A{D^{ - 1/\beta }}
		\label{DM_kernel}
	\end{equation}
where $A$ is the affinity matrix, and $D$ is the degree matrix of $A$. $\beta$ is the anisotropic diffusion parameter used by the diffusion operator, whose value is empirically set to 0.5~\citep{vos2020brainspace}. Then, the eigenvalue decomposition algorithm is used to extract the top $n$ largest eigenvalues ($\lambda$) along with their corresponding eigenvectors ($v$).
\begin{equation}
	{{P_\beta }v = \lambda v}
	\label{enginvalues}
\end{equation}
Finally, the diffusion embedding is constructed by scaling the eigenvectors to derive the $i$-th functional gradient component ${\Phi _i}$.
\begin{equation}
	{\Phi _i} = \frac{{{\lambda _i}}}{{1 - {\lambda _i}}} \cdot {v_i}
	\label{dm_4}
\end{equation}

	
The first $N_{gra}$ ($N_{gra}$=2) components of the functional gradients of each ROI were taken as input, and the unsupervised algorithm $K$-means was used to complete the clustering of ROIs. In this way, $N_R$ ROIs were clustered into $k$ ($k$=7) communities according to functional similarity. Then we customized a 1D convolution for each community to extract the ROIs features in this community as a representation of this community. The features extraction of each community can be summarized in the following formula. 
	\begin{equation}
		{C_i} = \sigma \left( {LN\left( {conv1{D_i}\left( {ROI{s_i}} \right)} \right)} \right)
		\label{conv1d}
	\end{equation}
where $ROI{s_i}$, $LN$, and $\sigma$ represent the feature matrix of ROIs belonging to the $i$-th community, layernorm, and activation function, respectively. And $conv1{D_i}$ is a one-dimensional convolution operation specific to each community. The parameter settings  of the $conv1{D_i}$ are listed in Table~\ref{block details}.


\begin{table*}[]
	\caption{Detailed architecture and parameters of community scale and global representation scale.}
	\label{block details}
	\resizebox{\textwidth}{!}{
	\begin{tabular}{l|l|c|c|c}
		\hline
		Module                           & Block                       & Layer       & Output size    & Explanation                                                                 \\ \hline
		\multirow{5}{*}{Community Scale} & \multirow{5}{*}{Conv1D$_i$} & Input       & (N$_i$, D$_1$) & N$_i$ stands for the number of ROIs belonging to community i, D$_1$=64      \\ \cline{3-4}
		&                             & Conv1d      & (1, D$_1$)     & outchannel=1, kernel\_size=15, padding='same',stride=1                      \\ \cline{3-4}
		&                             & LayerNorm   & (1, D$_1$)     &                                                                             \\ \cline{3-4}
		&                             & Activation  & (1, D$_1$)     &                                                                             \\ \cline{3-4}
		&                             & Dropout     & (1, D$_1$)     & dropout rate=0.3                                                            \\ \cline{1-4}
		&                             & Concatenate & (k, D$_1$)     & The features of k communities are concatenated to form a community network  \\ \cline{1-4}
		\multirow{5}{*}{Global Scale}    & \multirow{3}{*}{Conv2D}     & unsqueeze   & (1, k, D$_1$)  &                                                                             \\ \cline{3-4}
		&                             & Conv2d      & (D$_2$, 1, 1)  & outchannel=16, kernelsize=(k, D$_1$), stride=1, D$_2$=16                    \\ \cline{3-4}
		&                             & squeeze     & (D$_2$)        &                                                                             \\ \cline{2-4}
		& \multirow{2}{*}{Linear}     & Linear      & (D$_2$)        & in features=D$_2$, out features=D$_2$                                       \\ \cline{3-4}
		&                             & Activation  & (D$_2$)        & At this point, we have obtained a feature representation of the whole brain \\ \hline
		\end{tabular}}
\end{table*}

\subsubsection{The global representation scale and classification}
By clustering functional gradients and extracting representations of each community, we obtained a functional community network with the shape $(k, D_1)$. To effectively characterize the brain at a whole-brain scale, we proposed a global scale representation module, which consists of a 2D convolutional layer (with a kernel size equal to the scale of the community network) and a linear layer to achieve whole-brain representation. The parameter details of this module are listed in Table~\ref{block details}.

In order to train the model serving as feature extractor, after the global scale representation module, we used a liner layer and a softmax layer to achieve the supervised classification task (i.e., training process). After the model was trained, we used it as feature extractor (denoted as $f(\cdot)$), the outputs of the global scale representation module were used as the subject's characteristic features. 

To implement the classification, the prototype learning was introduced, which achieves classification by learning a set of prototypes, representing the centers of different classes~\citep{qiu2024novel}. The prototype of the $i$-th class is computed as follows:
	\begin{equation}
		{P_i} = \frac{1}{K}\sum\nolimits_k {f\left( {F{C_{k,i}}} \right)} 
		\label{prototype}
	\end{equation}
where $P_i (i=1,2)$ is the prototype of class $i$, $K$ represents the number of samples in each class, and $F{C_{k,i}}$ corresponds to the FCN of the $k$-th subject in the $i$-th class.
	
\begin{figure*}
		\centering
		\includegraphics[width=\linewidth]{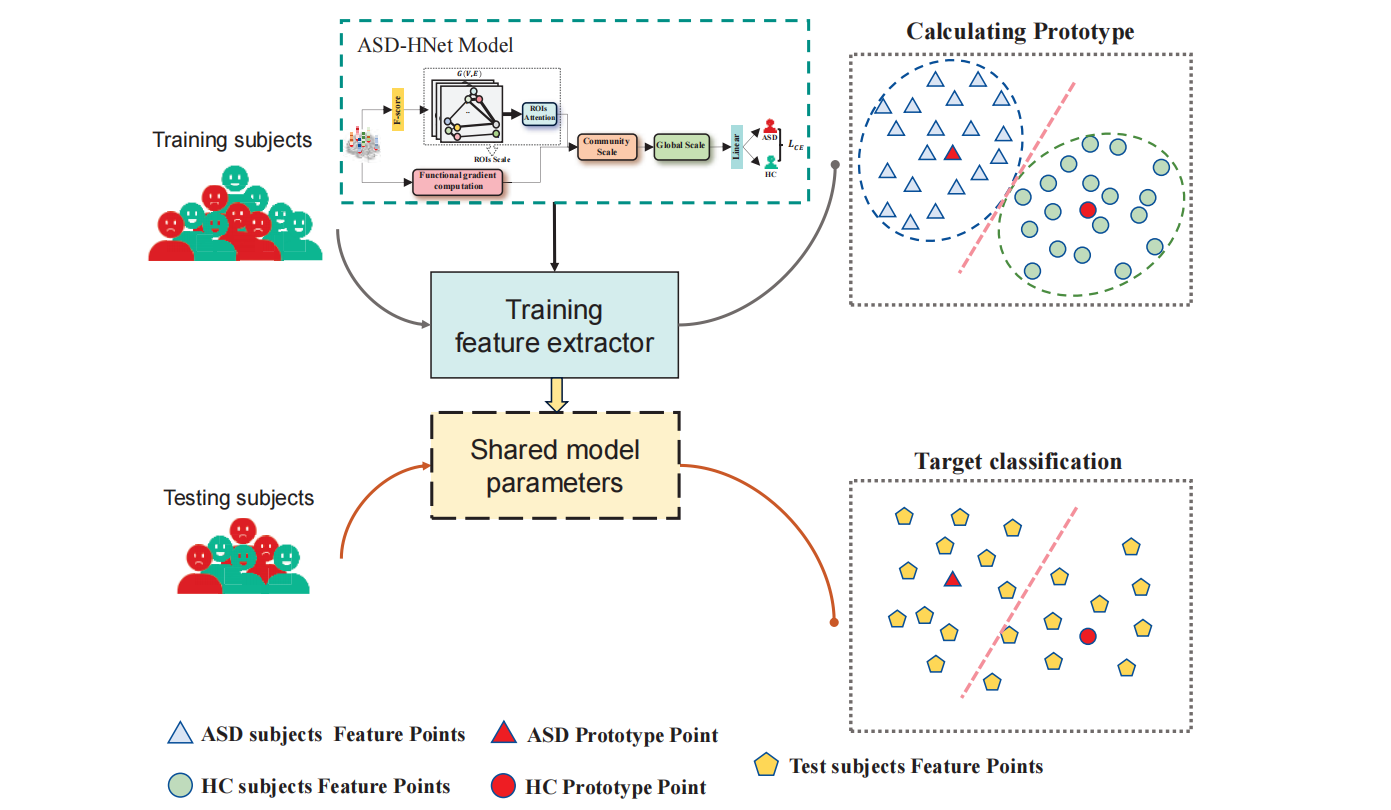}
		\caption{Workflow of prototype learning based classification. First, we use the training set to train the feature extractor, then calculate the corresponding prototype for each category. Finally, we obtain the features of the test data using the trained feature extractor and calculate the distance to each prototype  for final classification result.}
		\label{prototype_fig}
	\end{figure*}

For the final classification of test subjects, we calculated the Euclidean distance between the feature vectors of the test samples and the two prototypes. Each test sample was then assigned to the class of the prototype with minimal distance, as shown in Figure \ref{prototype_fig}. 
	
	\subsection{Compared methods}
	To demonstrate the performance of our proposed method, we compared ASD-HNet with state-of-the art methods, which are listed as follows.
	
	\subsubsection{f-GCN}
	F-GCN is a basic graph convolutional network. The focus of this method is to choose the appropriate threshold to remove redundant connections, with the optimal threshold identified in our experiment being 0.1. In f-GCN, multiple GCN layers are stacked. The model employs a global average pooling operator to generate a coarse graph, which reduces the number of parameters by decreasing the size of the representation and thus helps avoid overfitting. The readout layer then consolidates the node representations into a single graph representation~\citep{jiang2020hi}.
	
	\subsubsection{BrainnetCNN}
	Kawahara et al. proposed a new convolutional model named BrainnetCNN. BrainNetCNN incorporates specific convolutional and pooling layer designs to capture the complex relationships and features within brain networks. Typically, BrainNetCNN includes the following three main components, i.e., Edge-to-Edge (E2E) layer, Edge-to-Node (E2N) layer, Node-to-Graph (N2G) layer~\citep{kawahara2017brainnetcnn}. By combining these layers, BrainNetCNN effectively extracts useful features from brain network data.
	
	\subsubsection{CNN-model}
	Sherkatghanad et al. proposed a two-dimensional convolution of multiple different kernels to extract features at different levels, including features of single ROI and correlation features between ROIs~\citep{sherkatghanad2020automated}. And then the classification is achieved by fusing the features of different levels. In the end, the convolution with four different kernels performed best on processed data. For simplicity, we termed it as CNN-model, hereafter.
	
	\subsubsection{FBNETGEN}
	FBNETGEN is a learnable method to construct the adjacency matrix and explores the interpretability of the generated brain network for downstream tasks. Specifically, the method uses 1D-CNN and bi-GRU as encoders to learn the features of the time series, and then obtains the learned adjacency matrix by multiplying the learned features with its own transpose. Finally, graph convolution is used to implement the final classification task~\citep{kan2022fbnetgen}.
	
	\subsubsection{ASD-DiagNet}
	ASD-DiagNet is a collaborative learning strategy that combines single-layer perceptron and autoencoder. For the input data, a linear data enhancement approach is used to augment the training data for better feature extraction~\citep{eslami2019asd}. 
	
	\subsubsection{BrainTransformerNet}
	BrainTransformerNet leverages transformer to learn the edge attention of functional connectivity and extract an interpretable functional network. In this approach, a new readout function is also designed to take advantage of modular level similarities between ROIs in the brain network for better brain representation~\citep{kan2022brain}.
	
	\subsubsection{Hi-GCN}
	Hi-GCN takes the features of each subject extracted by f-GCN as nodes, and gets the edges between subjects through the calculation of correlation coefficients. After that, the population graph networks is built. Then, four-layer residual graph convolution is used to achieve feature transfer between subjects, and the learned feature representation of each subject is obtained to realize the final classification task~\citep{jiang2020hi}.
	
	\subsubsection{MCG-Net}
	Luo et al. proposed a dual-branch network architecture, MCG-Net. This method employs parallel convolution and graph convolution to respectively extract high-order representations and low-order spatial features of functional connectivity networks. Ultimately, the extracted multi-features are fused and utilized in a downstream classification task through linear layers.~\citep{luo2023aided}.

\subsection{Experimental setting and evaluation metrics}
In our experiment, we employed the 10-fold cross-validation strategy to conduct the comparison among different models. The data were divided into 10 folds, with 9 folds used for training and 1 fold used for testing. The models were implemented with PyTorch framwork, and all the experiments were conducted on a GeForce RTX 3090 GPU. 

We adopted the adaptive moment estimation optimizer (Adam) to train the model with using cross-entropy as the loss function. More hyperparameter settings in the model are listed in Table~\ref{hyperparameter setting}. The cross entropy loss function is defined as below:
\begin{equation}
	{L_{CE}} = \frac{1}{N}\sum\limits_i^N { - [{y_i} \times \log ({p_i}) + \left( {1 - {y_i}} \right) \times \log \left( {1 - {p_i}} \right)]}
	\label{crossentropy}
\end{equation}
where $N$ denotes the number of training subjects, $y_i$ is the true label of the $i$-th subject, and $p_i$ stands the predicted probability.

\begin{table}[]
	\caption{Hyper-parameters setting for the proposed model.}
	\label{hyperparameter setting}
	\begin{tabular}{llcrl}
		\hline
		& Hyperparameter      & \textbf{} & Setting      &  \\ \hline
		1 & Optimizer           &           & Adam         &  \\
		2 & Learning rate       &           & 0.0001       &  \\
		3 & Dropout rate        &           & 0.3          &  \\
		4 & Training batch size &           & 32           &  \\
		5 & Activation function &           & Tanh($\cdot$)         &  \\
		6 & Loss                &           & CrossEntropy &  \\ \hline
	\end{tabular}
\end{table}

For the hyperparameters $N_{gra}$ and $k$, we set them to be 2 and 7 during the experiments, respectively. In subsection~\ref{hyperparameter_section}, we will discuss the influence of different settings on the classification performance. 

In order to evaluate the model, four metrics, i.e., sensitivity (Sen), specificity (Spe), accuracy (Acc), and F1, were applied as evaluation indicators. The final values of these metrics were averaged across ten results from 10-fold cross-validation. 
The definitions were as follows:

\begin{equation}
	Acc = \frac{{TP + TN}}{{TP + FN + FP + TN}}
	\label{ACC}
\end{equation}

\begin{equation}
	Sen = \frac{{TP}}{{TP + FN}}
	\label{SEN}
\end{equation}

\begin{equation}
	Spe = \frac{{TN}}{{TN + FP}}
	\label{SPEC}
\end{equation}

\begin{equation}
	F1 = \frac{{2TP}}{{2TP + FP + FN}}
	\label{F1}
\end{equation}
where TP, FN, TN and FP represent the number of ASD subjects correctly classified, the number of ASD subjects wrongly classified as HC subjects, the number of HC subjects correctly classified, the number of HC subjects wrongly classified as ASD subjects, respectively.

\section{Results and discussion}
\subsection{Performance comparison among different methods}
The classification results of the ASD-HNet and compared methods are shown in Table~\ref{comparison}, with the same experimental settings. All the results of the compared methods were obtained by reproducing the models according to the public codes provided in the references. The hyper-parameters were set according to the original studies, except for f-GCN and Hi-GCN. During training the f-GCN and Hi-GCN, we set the corresponding optimal threshold to be 0.1 and 0.2 respectively. It can be found that the proposed ASD-HNet outperforms other methods on all four indicators, with accuracy improved by 3.09\%, sensitivity improved by 2.94\%, specificity improved by 1.38\% and F1-score  improved by 2.45\%, respectively. To further validate the effectiveness of the proposed method, we performed a paired t-test on the ten-fold cross-validation accuracy of the models. The results reveal that the p-values for comparisons between ASD-HNet and all other methods are less than 0.05, demonstrating the statistical significance of the performance improvements achieved by the proposed method.

\begin{table*}[!htbp]
		\caption{Performance comparison of among the proposed models and other models. The best results are marked in bold. The results marked with an asterisk (*) are reproduced.}
		\label{comparison}
		\begin{tabular}{lccccc}
			\hline
			Method               & Accuracy(\%)   & Sensitivity(\%) & Specificity(\%) & F1-score(\%)   & \multicolumn{1}{l}{p-value} \\ \hline
			f-GCN*               & 61.32          & 61.28           & 61.31           & 59.13          & 0.0002                      \\
			BrainnetCNN*         & 64.52          & 59.84           & 68.46           & 61.03          & 0.0006                      \\
			CNN-model*           & 64.88          & 58.55           & 69.70           & 60.36          & 0.0053                      \\
			FBNETGEN*            & 67.14          & 69.69           & 64.65           & 68.59          & 0.0029                      \\
			ASD-DiagNet*         & 66.70          & 60.64           & 71.91           & 62.56          & 0.0023                      \\
			BrainTransformerNet* & 67.00          & 64.23           & 70.23           & 67.14          & 0.0151                      \\
			Hi-GCN*              & 67.34          & 68.78           & 65.70           & 65.17          & 0.0052                      \\
			MCG-Net*             & 70.48          & 67.08           & 73.52           & 68.32          & 0.0375                      \\
			ASD-HNet             & \textbf{73.57} & \textbf{72.63}  & \textbf{74.90}  & \textbf{71.04} & -                           \\ \hline
		\end{tabular}
	\end{table*}

To further evaluate the performance of ASD-HNet, we also compared our ASD-HNet method to other four methods that performed well on the ABIDE dataset, i.e., MC-NFE~\citep{wang2022multi}, MVS-GCN~\citep{wen2022mvs}, MBT~\citep{dong2024multiview} and DGAIB~\citep{dong2024brain}. In these studies, the authors adopted different amounts of data and different brain templates. Namely, MC-NFE used a total of 609 subjects recruited from 21 sites in ABIDE-I and ABIDE-II~\citep{wang2022multi}. The MVS-GCN used 871 subjects from 17 different sites, including 403 patients with ASD and 468 normal subjects~\citep{wen2022mvs}.  MBT chose 1035 subjects, including 505 patients diagnosed with ASD and 530 healthy controls~\citep{dong2024multiview}. DGAIB pooled 1096 subjects from 17 sites in ABIDE-I, consisting of 569 healthy controls and 527 individuals diagnosed with ASD~\citep{dong2024brain}.

\begin{table*}[!htbp]
		\caption{Comparison with state-of-the-art methods(ABIDE), with best results shown in bold. '-' indicates the experiment results are not reported on the dataset.}
		\label{sota-compare}
		\begin{tabular}{lccccc}
			\hline
			\begin{tabular}[c]{@{}c@{}}Method\\ (ASD/HC)\end{tabular} & \begin{tabular}[c]{@{}c@{}}MC-NFE\\ \citep{wang2022multi}\\ (280/329)\end{tabular} & \begin{tabular}[c]{@{}c@{}}MVS-GCN\\ \citep{wen2022mvs}\\ (403/468)\end{tabular} & \begin{tabular}[c]{@{}c@{}}MBT\\ \citep{dong2024multiview}\\ (505/530)\end{tabular} & \begin{tabular}[c]{@{}c@{}}DGAIB\\ \citep{dong2024brain}\\ (527/569)\end{tabular} & \begin{tabular}[c]{@{}c@{}}ASD-HNet\\ (Ours)\\ (396/452)\end{tabular} \\ \hline
			Atlas          & BASC-064        & AAL-116         & AAL-116         & AAL-90           & AAL-116         \\ 
			ACC(\%)        & 68.42           & 68.92           & 72.08           & 71.20            & \textbf{73.57}    \\ 
			SEN(\%)        & 70.05           & 69.09           & 72.26           & 69.60            & \textbf{72.63}    \\ 
			SPE(\%)        & 63.64           & 63.15           & 71.50           & 73.41            & \textbf{74.90}    \\ 
			F1-score(\%)   & -               & -               & \textbf{71.69}           & 71.39   & 71.04           \\ \hline
		\end{tabular}
\end{table*}
	
For a intuitive comparison between these method and our ASD-HNet, we directly quoted the reported results from original references as previous studies~\citep{wang2022multi}. In these studies, the authors did not release the implementation codes, or did not describe how to generate the used data from ABIDE dataset. The comparison results are summarized in Table~\ref{sota-compare}. By this rough comparison, we can also find that although the F1-score of our method is little lower than MBT and DGAIB methods, our method achieved the best accuracy (73.57\%), sensitivity (72.63\%) and specificity (74.90\%).
	
\subsection{Ablation experiment}
To evaluate the rationality and effectiveness of the proposed model's architecture, we conducted a series of ablation experiments to assess the impact of removing each key module on overall performance. Specifically, to validate the effectiveness of information transfer in the ROI-scale graph convolution and to determine whether the ROIs attention module contributes to improved model performance, we performed comparative experiments with and without the graph convolutional network and the ROIs attention mechanism. Additionally, to test the validity of prototype learning, we trained the feature extractor using only the original linear layer for classification, thereby isolating the contribution of prototype learning to the model's performance. The results of the ablation experiment were shown in Table~\ref{ablation}.   

	\begin{table*}[!htbp]
		\caption{The results of ablation experiments. The best results are marked in bold.}
		\label{ablation}
		\begin{tabular}{lcccc}
			\hline
			Method                                 & Accuracy(\%)   & Sensitivity(\%) & Specificity(\%) & F1-score(\%)   \\ \hline
			w\textbackslash{}o ROI-scale GCN       & 70.12          & 67.58           & 72.51           & 68.13          \\
			w\textbackslash{}o ROIs attention      & 71.79          & 69.85           & 73.53           & 70.04          \\
			w\textbackslash{}o Prototype Leanrning & 72.86          & 67.83           & \textbf{77.07}  & 69.82          \\
			ASD-HNet                                & \textbf{73.57} & \textbf{72.63}  & 74.90           & \textbf{71.04} \\ \hline
		\end{tabular}
	\end{table*}
	
Through the ablation experiment results, we can find that the performance of the model decreased significantly after the elimination of three modules respectively. Although removing prototype learning decreased the specificity, the proposed model could provide higher sensitivity for effectively diagnosing the ASD subjects, and achieved more balanced performance on the two metrics of sensitivity and specificity.
	
\subsection{The influence of the hyperparameters settings on the model performance}
\subsubsection{Influence of the $N_{gra}$ and $k$}~\label{hyperparameter_section}
\begin{figure*}
		\centering
		\includegraphics[width=\linewidth]{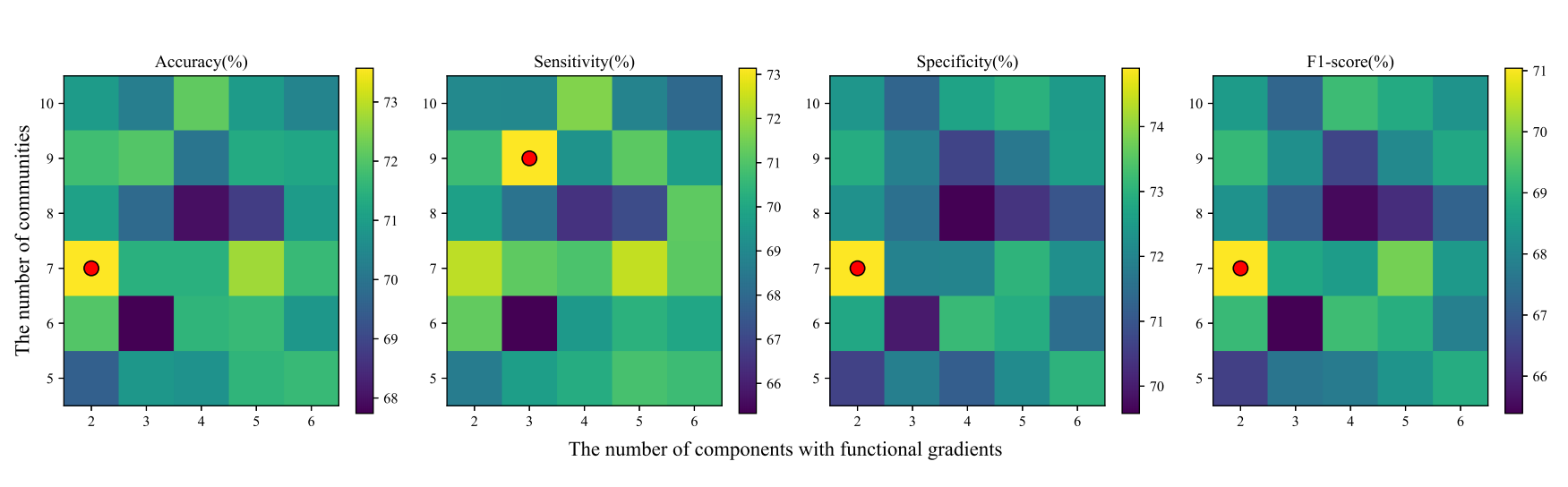}
		\caption{The averaged four metrics obtained by different combination of the parameters $k$ and $N_{gra}$.}
		\label{Search-hyperparameters}
	\end{figure*}

In ASD-HNet, two hyperparameters $N_{gra}$ and $k$ need to be configured, which represent the number of functional gradients components and the number of functional communities, respectively. The results reported above were obtained when $N_{gra}$ and $k$ were set to be 2 and 7. Different combinations of these two hyperparameters may lead to varying classification performance. To investigate the impact of hyperparameter settings on model performance, we conducted the systematic hyperparameter search experiments as previous studies~\citep{jiang2020hi,wen2022mvs,wang2024leveraging}. We conducted experiments with $k$ ranging from 5 to 10 and $N_{gra}$ ranging from 2 to 6. The experimental results are illustrated in Figure~\ref{Search-hyperparameters}. The highest score for each metric is marked with a red circle.

From the results in the Figure~\ref{Search-hyperparameters}, we can find that although the best combination of hyperparameter values in the sensitivity index is different from the other three indicators, the three indicators of accuracy, specificity and F1-score all achieve the best performance when $k$=7 and $N_{gra}$=2. Moreover, sensitivity has the third highest score under this combination of hyperparameter values. In previous study, Gong et al. found that the first two functional gradients can explain the majority of variance in functional networks~\citep{gong2023connectivity}. Similarly, Samara et al. also found similar results in their research~\citep{samara2023cortical}. Our experimental results are consistent with previous findings. Based on comprehensive evaluation, we set the $k$ and $N_{gra}$ as 7 and 2 in all our experiments.

\subsubsection{Influence of the Node Feature Retention Ratio}
\begin{figure}
	\centering
	\includegraphics[width=\linewidth]{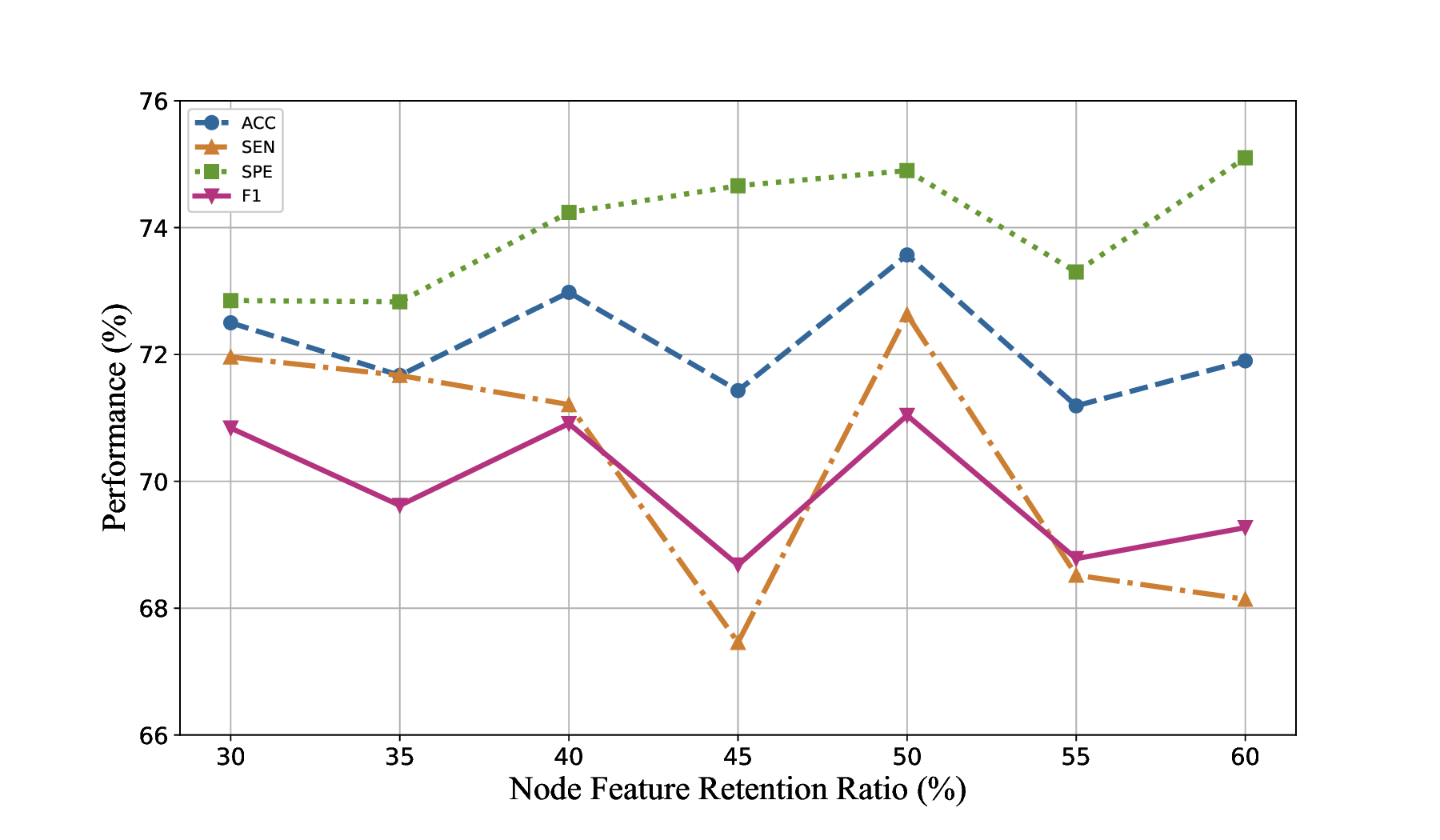}
	\caption{Impact of node feature retention ratio (${\lambda _R}$) on model performance.}
	\label{node-ratio}
\end{figure}

In the ROIs scale, there exists a hyperparameter ${\lambda _R}$ (node feature retention ratio). To investigate the impact of this parameter on the overall structural performance, we conducted a hyperparameter search experiment. The ${\lambda _R}$ ranges from 30\% to 60\% with a step size of 5\%. The experimental results are presented in Figure~\ref{node-ratio}. It can be found that when the retention ratio is set at 50\%, the model achieves optimal performance in terms of accuracy, sensitivity, and F1-score, while the specificity metric is only slightly lower than that at the 60\% ratio. An appropriate retention ratio enables the model to better learn useful features. When the retention ratio is too low, some features beneficial for classification may be inadvertently removed. Conversely, when the retention ratio is too high, an excessive number of noisy features may be retained. Both scenarios can lead to suboptimal model performance. Therefore, we set ${\lambda _R}$ to be 50\% in current study to report the experimental results.

\subsection{The interpretability and visualization analysis}

In this section, we explore the interpretability of the proposed method and the learned ROIs attention, investigating whether the discriminative features identified are interpretable and consistent with previous findings.

\begin{figure}[h]
		\centering
		\includegraphics[width=\linewidth]{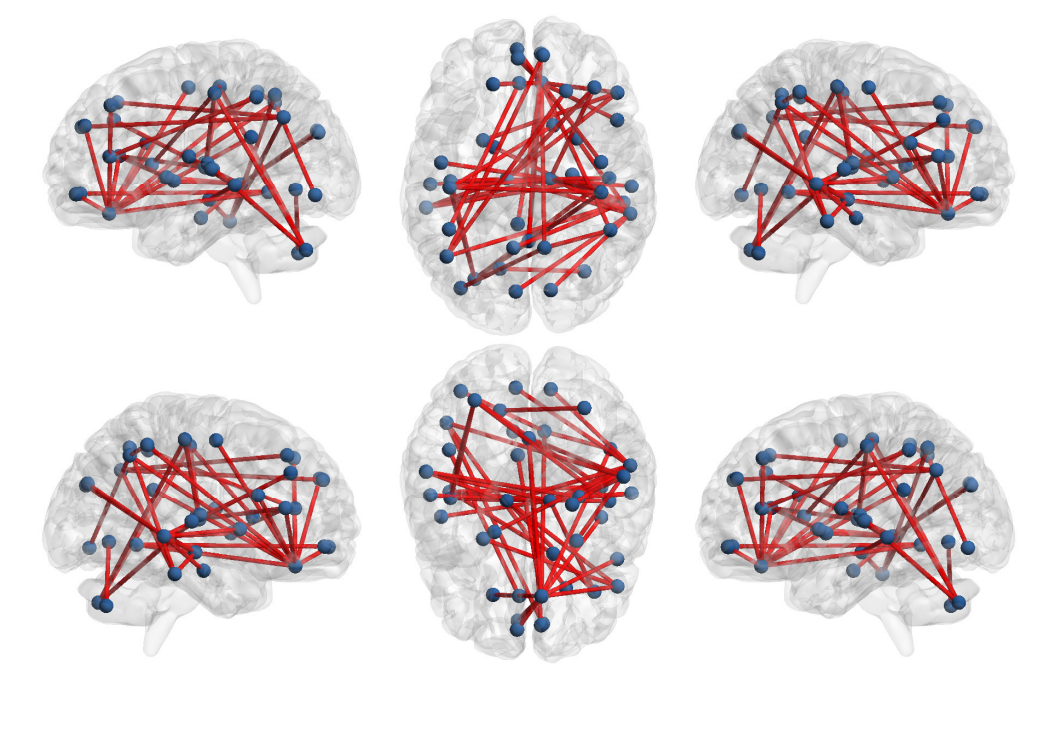}
		\caption{The edges of the brain regions with the greatest difference between HC and ASD patients.}
		\label{discriminal-edges}
	\end{figure}

First, we want to check whether the model can find the difference patterns of FCN between ASD subjects and HCs subjects. F-score is a technique for extracting key features based on a data-driven approach. For the analysis of key edges of brain regions, we retained the $10\%$ edges with the highest F-score for retention and visualization, as shown in Figure~\ref{discriminal-edges}. All the visualization are achieved with BrainNet Viewer~\citep{xia2013brainnet} tool. The most differentiated edges observed in Figure~\ref{discriminal-edges} are mostly the connections between the temporal lobe, the frontal lobe and the sensorimotor area, which is consistent with previous findings. For instance, Abrams et al. has reported an atypical decline in functional connectivity between the temporal area (auditory network) and the medial prefrontal area (DMN) in people with autism~\citep{abrams2013underconnectivity}.
	
Furthermore, we want to show that the functional gradients in the model can find the difference patterns on the community networks. We conducted the analysis of the validity and interpretability of utilizing the first two components of functional gradients for constructing community networks. Since the gradients calculated separately for the two groups of levels may not be directly comparable~\citep{vos2020brainspace}, we chose the joint embedding~\citep{xu2020cross} to realize the alignment of the two groups of functional gradients. Then we conducted a joint analysis of the principal and secondary gradients.

\begin{figure*}
		\centering
		\begin{subfigure}[b]{0.45\textwidth}
			\centering
			\includegraphics[width=\textwidth]{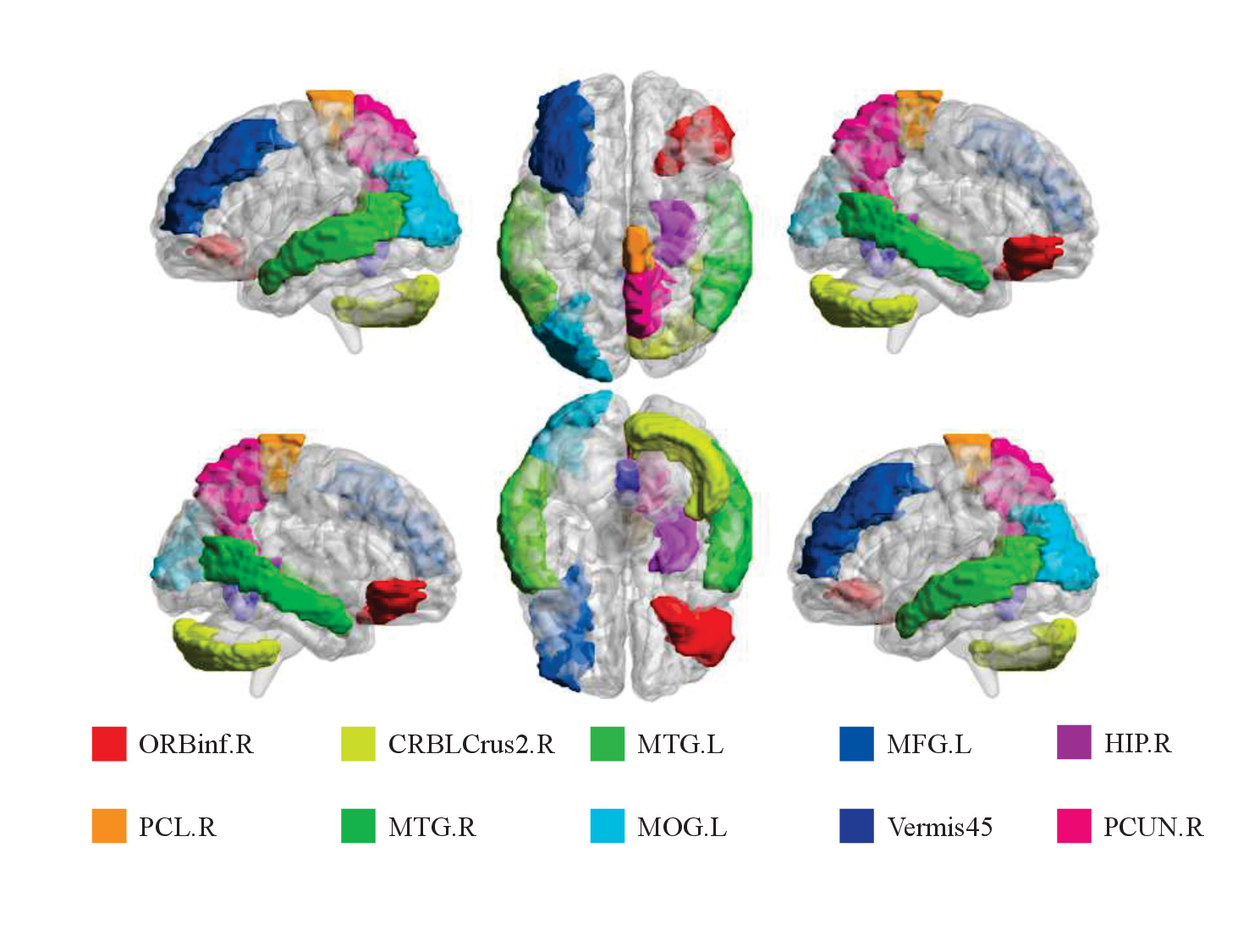}
			\caption{The ten ROIs with the greatest differences in the first two gradients.}
			\label{fig8-1}
		\end{subfigure}
		\hspace{0.015\textwidth}
		 \begin{subfigure}[b]{0.45\textwidth}
			\centering
			\includegraphics[width=\textwidth]{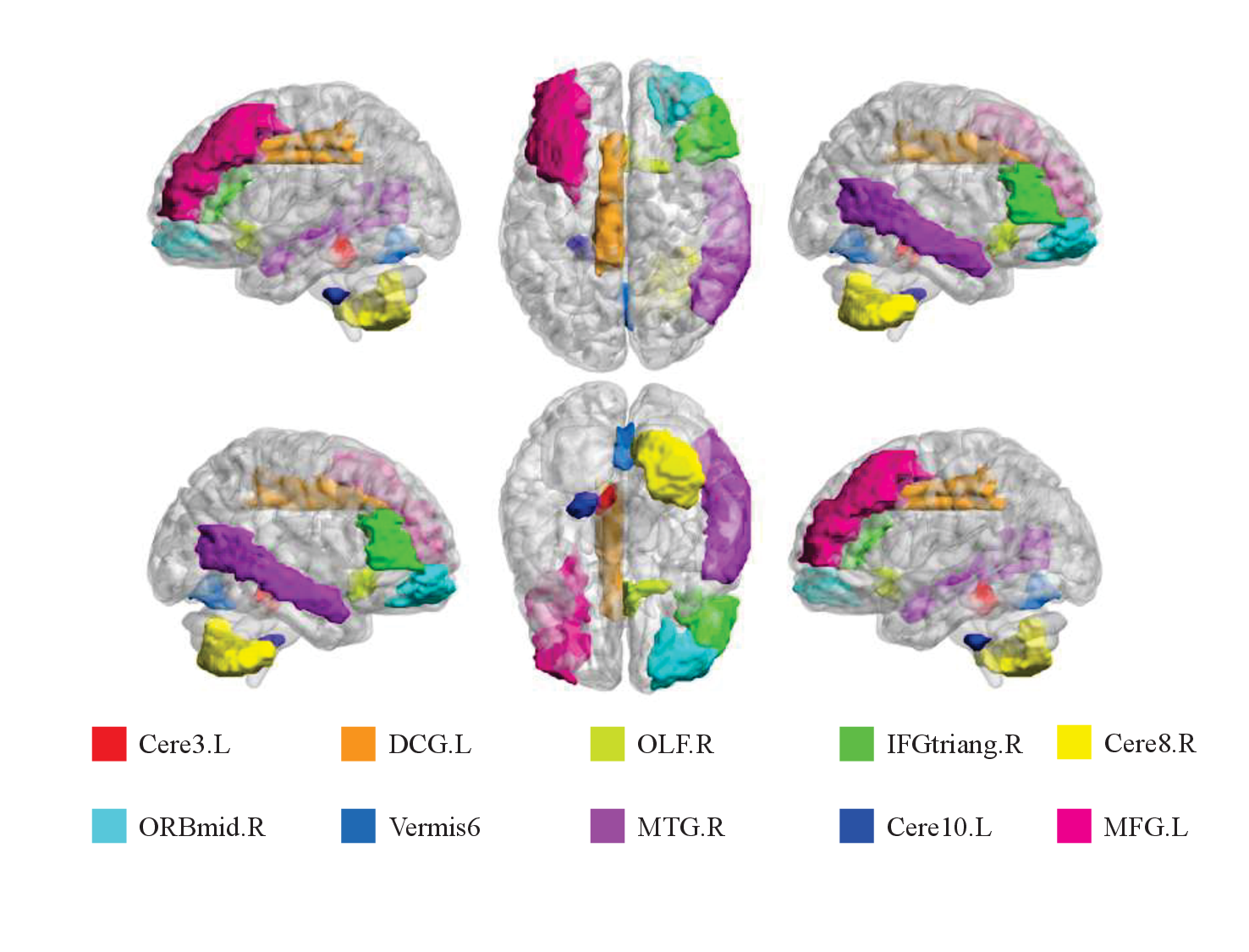}
			\caption{The ten ROIs with the highest attention scores obtained from ROIs attention scores.}
			\label{fig8-2}
		\end{subfigure}
		\caption{The ten ROIs with the greatest differences in the first two functional gradients and the ten ROIs with the most significant differences learned by ROIs attention weights.}
	\end{figure*}

We merged the first two gradients (principal and secondary gradients) and employed Euclidean distance (consistent with K-Means algorithm) to measure the distinction in corresponding ROIs between the ASD and HC groups. Subsequently, we visualized the ten brain regions with the greatest Euclidean distance (indicating the most significant difference). The visualized result is shown in Figure \ref{fig8-1}.
	
Based on the aforementioned findings, we can observe that the group-level functional gradient differences are primarily reflected in the following brain regions: the inferior orbital gyrus (ORBinf.R), right cerebellar Crus II (CRBLCrus2.R), left middle temporal gyrus (MTG.L), left middle frontal gyrus (MFG.L), right hippocampus (HIP.R), right precuneus (PCL.R), right middle temporal gyrus (MTG.R), left middle occipital gyrus (MOG.L), vermis (Vermis45), and right precuneus (PCUN.R). The MTG has been widely confirmed to be associated with ASD~\citep{xu2020specific}. And the cerebellum, due to its crucial role in regulating brain function, has significant associations with ASD~\citep{mosconi2015role, hampson2015autism}. Xiao et al. also confirmed the association between the precuneus and ASD~\citep{xiao2023atypical}. These key ROIs and their associations are consistent with the findings of~\citep{hong2019atypical}.

Next, we visualize the ROIs attention weights learned by the model. The ten ROIs with the highest attention scores are visualized in Figure~\ref{fig8-2}. These ROIs include: left cerebellum (Cere3.L, Cere10.L), left dorsal cingulate gyrus (DCG.L), right olfactory cortex (OLF.R), right triangular part of the inferior frontal gyrus (IFGtriang.R), right cerebellum (Cere8.R), right middle orbital gyrus (ORBmid.R), vermis6, right middle temporal gyrus (MTG.R), and left middle frontal gyrus (MFG.L). We observe that several regions overlap between the group-level differences identified by the functional gradient and those highlighted by the ROIs attention scores, particularly in the temporal lobe~\citep{xu2020specific}, left prefrontal cortex~\citep{lee2009functional}, and certain cerebellar areas~\citep{mosconi2015role, hampson2015autism}.
	
	
	%
	
\subsection{Experiments on model generalization}
In order to investigate the generalization performance of the proposed method, we conducted experiments on both the ADHD-200 Consortium (ADHD-200) dataset~\footnote{\url{http://fcon_1000.projects.nitrc.org/indi/adhd200/}} and the ABIDE-II dataset~\citep{di2017enhancing}.

\subsubsection{ADHD-200 Dataset}
ADHD-200 dataset has 973 resting state fMRI acquisitions labeled with ADHD or healthy controls from subjects aging from 7 to 21. The dataset is collected from eight different international sites, which are NeuroImage group (NeuroImage), New York University Child Study Center (NYU), Peking University (Peking), Brown University (BU), Kennedy Krieger Institute (KKI), University of Pittsburgh (Pitts), Oregon Health and Science University (OHSU) and Washington University in St. Louis (WU).

We choose the dataset preprocessed by Athena pipeline~\citep{bellec2017neuro}. A total of 759 subjects from 7 sites were contained for fMRI data, including 303 ADHD patients and 456 HCs. Similar to the above experiments on ASD, we used 10-fold cross-validation to conduct corresponding experiments. Initially, for ASD-HNet, we used the same hyperparameters as we did for classifying ASD. We then tuned the model based on the specifics of the ADHD data. The adjustments were as follows: (1) Due to the imbalance in the ADHD data, we used a larger batchsize (64) to ensure that each batch included a more balanced representation of both classes. (2) Given the presence of multiple subtypes of ADHD \citep{molavi2020adhd}, we incorporated more global brain features from different perspectives to enhance the model's discriminative ability. Specifically, we increased the out$\_$channels of Conv2d in the global brain scale to 32. These adjustments can enhance the model's adaptation and optimization for classifying ADHD.
	 
In addition to the comparative methods mentioned above for ASD, we also included two state-of-the-art methods in our analysis: (1) MDCN \citep{yang2023deep}, which utilize fMRI data from 572 subjects, including 261 ADHD patients and 311 healthy controls (HCs); (2) A-GCL~\citep{zhang2023gcl}, which uses data from 947 subjects, comprising 362 ADHD patients and 585 HCs. The experimental results are shown in Table~\ref{ADHD-table}. The results indicate that our method also achieves better performance in classifying ADHD. Due to the imbalance in the data we used, the comparison methods we reproduced show a significant imbalance in specificity and sensitivity, resulting in a lower F1-score. However, our method achieves the best performance in terms of accuracy (71.33\%) and specificity (80.30\%). This demonstrated that ASD-HNet not only performs well in classifying ASD but also achieved good classification performance for ADHD, which validates its robustness and generalization.

	\begin{table*}
		\caption{Experimental results of ASD-HNet on ADHD-200 data, with best results shown in bold. '*' stands that the method has not been reproduced, and the results shown are those reported in the original study and '-' indicates the experiment results are not reported on the dataset. "Tuned" denotes the adjustments made to ASD-HNet through fine-tuning.}
		\begin{tabular}{llcccc}
			\hline
			ADHD-200                                                     & Method                                                           & ACC(\%)        & SEN(\%)        & SPE(\%)        & F1-score(\%)   \\ \hline
			& f-GCN                                                            & 61.12          & \textbf{83.47} & 27.43          & 30.95          \\
			& BrainnetCNN                                                      & 62.53          & 49.64          & 71.84          & 51.01          \\
			& CNN-model                                                        & 64.13          & 43.69          & 77.64          & 49.33          \\
			& Hi-GCN                                                           & 66.01          & 78.96          & 46.58          & 51.82          \\
			& MCG-Net                                                          & 68.67          & 53.30          & 79.41          & 57.26          \\ \hline
			\begin{tabular}[c]{@{}l@{}}(ADHD/HC)\end{tabular} & \begin{tabular}[c]{@{}l@{}}MDCN* \\ (261/311)\end{tabular}  & 67.45          & 71.97          & 62.39          & \textbf{66.86} \\
			& \begin{tabular}[c]{@{}l@{}}A-GCL* \\ (362/585)\end{tabular} & 70.92          & -              & -              & -              \\ \hline
			Ours                                                         & ASD-HNet                                                          & 70.00          & 57.22          & 78.53          & 60.44          \\
			& \begin{tabular}[c]{@{}l@{}}ASD-HNet (tuned)\end{tabular}        & \textbf{71.33} & 58.49          & \textbf{80.30} & 61.80          \\ \hline
		\end{tabular}
		\label{ADHD-table}
	\end{table*}

\subsubsection{ABIDE-II Dataset}
The ABIDE-II initiative was established as an extension of ABIDE-I, aiming to further advance scientific research on the brain connectome in Autism Spectrum Disorder. ABIDE-II has aggregated over 1000 additional datasets, offering more comprehensive phenotypic characterizations, particularly in the quantitative assessment of core ASD symptoms and associated features. The project now encompasses 19 research sites, including 10 founding institutions and 7 newly participating members, collectively contributing 1114 datasets from 521 individuals with ASD and 593 healthy controls (age range: 5–64 years). These data were openly released to the global scientific community in June 2016, providing a critical resource for ASD research.

In addition to the ADHD-200 dataset, we also evaluated the generalization performance of ASD-HNet on the ABIDE-II dataset, which consists of fMRI data from individuals with autism spectrum disorder and matched healthy controls. For this dataset, we used data from five sites, i.e., BN, EMC, GU, NYU, and OHSU, including a total of 382 subjects, comprising 185 ASD patients and 197 healthy controls. The preprocessing procedure for this dataset followed the same pipeline as ABIDE-I. Since the sample size of this dataset is smaller than that of the ABIDE-I and ADHD-200 datasets, we employed 5-fold cross-validation to balance training efficiency and validation accuracy. The experimental results are presented in Table~\ref{ABIDE-II}, five compared methods were selected as the baselines.


The experimental results demonstrate that the ASD-HNet method achieves the best performance, which achieves performance improvements of 4.43\% in accuracy, 7.47\% in specificity, and 2.32\% in F1-score compared to the the second-best method. These additional results further validating the generalization ability of the proposed method across different datasets for the same disease.
\begin{table*}[]
	\caption{Experimental results of ASD-HNet and compared methods on ABIDE-II data, with the best results are highlighted in bold.}
	\begin{tabular}{llcccc}
		\hline
		ABIDE-II & Method         & ACC(\%)        & SEN(\%)        & SPE(\%)        & F1-score(\%)   \\ \hline
		& f-GCN          & 62.40          & 56.20          & 60.30          & 52.33          \\
		& BrainnetCNN    & 68.12          & \textbf{69.72} & 62.12          & 66.43          \\
		& CNN-model      & 66.85          & 59.21          & 70.57          & 62.49          \\
		& Hi-GCN         & 64.17          & 64.17          & 56.58          & 56.63          \\
		& MCG-Net        & 65.51          & 61.55          & 64.06          & 62.00          \\
		& ASD-HNet(ours) & \textbf{72.55} & 65.23          & \textbf{78.04} & \textbf{68.75} \\ \hline
	\end{tabular}
	\label{ABIDE-II}
\end{table*}

\subsection{Limitation}
Even though ASD-HNet achieved better performance than other methods, some limitations still exist in this study. First of all, the classification performance is still unsatisfactory and can not be applied to clinical scenes in current stage. External cohort datasets should be collected for further validation~\citep{HAN201944, Wang2021Atypical}. The proposed method relies on a large amount of data for corresponding training. When the amount of data is insufficient or unbalanced, how to maintain the performance of the model is an important issue worthy of further research. Besides, we did not consider the large variability among the dataset from different sites into the model construction in current study. It is beneficial to explicitly eliminate the differences on data distribution with some technologies, such as domain adaption~\citep{wang2019identifying, wen2024training, liu2023domain}. We have also recognized the issue of site heterogeneity and have conducted preliminary experiments to mitigate its impact, as discussed in~\citep{luo2025multisitersfmridomainalignment}. In the future, we need address these issues during designing models to conduct the multi-site fMRI data analysis.
	
\section{Conclusion}
In this work, we proposed a hybrid neural network model to hierarchically extract features on the functional brain networks for ASD identification. Extensive experiments demonstrate the effectiveness of the proposed model. Compared with other baseline methods, the proposed model achieve superior performance on widely used dataset ABIDE-I. The ablation experiments and the interpretability analysis further verify the rationality of the model architecture. The generalization performance was also explored preliminarily on the ADHD-200 dataset and ABIDE-II dataset. In future studies, it is beneficial to integrate neural mechanisms into the deep learning models to boost the ASD identification.

\section*{Acknowledgments}
This work was supported in part by the National Natural Science Foundation of China under Grant No.62076209.

	
	\printcredits
	\bibliographystyle{cas-model2-names}
	\bibliography{reference}
	
\end{sloppypar}
\end{document}